# Can the conductance of an adiabatic ballistic constriction be lower than $2e^2/h$?

C.-T. Liang
*Department of Physics, National Taiwan University, Taipei 106, Taiwan*

O. A. Tkachenko, V. A. Tkachenko, D. G. Baksheyev
*Institute of Semiconductor Physics, Siberian Division, Russian Academy of Sciences, Novosibirsk, 630090 Russia*

M. Y. Simmons
*School of Physics, University of New South Wales, Sydney 2052, Australia*

D. A. Ritchie, M. Pepper
*Cavendish Laboratory, Madingley Road, Cambridge CB3 0HE, United Kingdom*

We have performed four-terminal conductance measurements of a one-dimensional (1D) channel in which it is possible to modulate the potential profile using three overlaying finger gates. In such a 1D ballistic structure we have observed, *for the first time,* that the conductance steps show a gradual decrease from $2e^2/h$ to $0.97 \times 2e^2/h$ with increasing negative finger gate voltage in a short, clean 1D constriction. We suggest this phenomenon is due to differing shifts of 1D subbands with changing spilt-gate voltage. Both a simple analytical estimate for an adiabatic constriction and, realistic modeling of the device, give the same magnitude of the conductance decrease as observed in our experiments.

It is known that the conductance of a one-dimensional (1D) ballistic wire is quantised in units of $2e^2/h$. This was discovered experimentally [1] in split-gate induced constrictions in high-mobility two-dimensional electron gases (2DEGs). Both experimental [2] and theoretical results [3] have shown that the accuracy of the observed quantisation is sensitive to the detailed shape of the confining potential and the presence of impurities. The observation that a high number of conductance plateaux [4] are exactly quantised, without the appearance of any resonant features indicates that adiabatic transport in the constriction [5,6] is attainable. However, regular deviations of plateaux from quantised values $n \times 2e^2/h$ have been observed in long 1D wires [7–10] and have been associated with either nonadiabaticity of the constriction or electron-electron interactions in the 1D system. In the seminal work of Tarucha, Honda and Saku [7] a few percent deviations were observed in long (5–10 $\mu$m) weakly disordered quantum wires [7]. The deviation was observed to increase with increasing wire length and was explained in terms of electron-electron interactions in a "dirty Luttinger liquid" [11]. The reduction of conductance plateaux of up to 25% has also been observed in 5–6 $\mu$m quantum wires and was ascribed to backscattering of electrons due to impurities [10]. Other structures with a more complicated geometry (T-shaped cleaved-edge-overgrowth quantum wires [8] and V-groove quantum wires [9]) have shown a significant decrease of the value of the conductance steps and have been attributed to backscattering of electrons from the abrupt interface between the 2D contacts and the 1D wire.

In this paper, we show *for the first time,* that a deviation from the quantised value $2e^2/h$ is possible even in a clean, short constriction in which residual scattering and effects of non-adiabaticity are negligible. It is well established [12] that the two-terminal conductance of a clean ballistic 1D system is determined by non-interacting source and drain contacts. In such a clean system, deviations from exact quantisation arising from the presence of electron-electron interactions are not expected [12], since these only occur in weakly disordered quantum wires ("dirty Luttinger liquid") [11]. Therefore the observed deviation in our system could not be explained by conductance renormalisation due to electron-electron interactions.

We have studied constrictions electrostatically shaped in the 2DEG by a pair of split gates and three finger gates lying across the channel. This versatile device permits control of the electrostatic potential across and along the channel without changing the series resistance, $R_s$. We have measured the conductance $G$ as a function of split gate voltage $V_g$ at different centre finger gate voltages, $V_F$ with outer finger gates grounded to the 2DEG. With increasing negative voltage on the centre finger gate we observed in a series of $G(V_g)$ curves the transition from pronounced steps to smeared conductance steps. At the same time the height of the first conductance step gradually decreased from $2e^2/h$ to $0.97 \times 2e^2/h$. The effect was reproduced in two- and four-terminal conductance measurements, whereas the series resistance $R_s = R(V_g{=}0)$ is varied by an order of magnitude. We ascribe this deviation from the conductance quantum $2e^2/h$ to the interplay between reflection in the first transmission mode and tunneling via the second mode in the adiabatic constriction. Electrostatics calculations show that the tops of the first and second subbands move with different velocities in response to the change of the split-gate voltage. This leads to a slower decrease of the reflection as compared to simultaneous increase in the tunneling, and therefore onset of the conductance step occurs at a lower value than $G = 2e^2/h$. The effect disappears when the smeared conductance steps turn into pronounced ones.

The multi-layered gated 1D structure was lithographically defined 158 nm above the 2DEG as shown in Fig. 1 (a). There is a 30-nm-thick layer of polymethylmethacrylate (PMMA) which has been highly irradiated by an electron beam, to act as a dielectric between the split-gate and three overlaying finger gates. The 2DEG has a carrier density of $2.5 \times 10^{15}$ cm$^{-2}$ with a mobility of $3.0 \times 10^6$ cm$^2$/Vs after brief illumination with a red light emitting diode. Experiments were performed in a pumped $^3$He cryostat at 300 mK and the four-terminal resistance was measured using an ac driving current of 10 nA at a frequency of 77 Hz with standard phase-sensitive techniques.

In our set-up, the outer finger gates were grounded to the 2DEG, while the centre finger gate voltage, $V_F$ and the split gate voltage, $V_g$ were varied over a wide range. In this sample, potentiometric contacts were deliberately brought close to the constriction (four-terminal measurements) to reduce the contribution of series resistance of the 2DEG (Fig. 1(b)) The measured resistance $R_s$ between the contacts in the absence of the constriction was only 115 Ω, that is 0.9% of the quantum resistance $h/2e^2$. Thus, we were able to measure the conductance of the single-mode constriction precisely enough without resorting to subtraction of the series resistance (Fig. 1(c)). Notice that the series resistance was 725 Ω in the two-terminal measurements.

Figure 2 shows the measured dependencies of conductance $G$ on (a) the split gate voltage $V_g$ and (b) the derivative $dV_g(G)/dG$ for different finger gate voltages from 0 to -4.8 V. If we consider the first conductance step we can determine its height $G_*$ by the value of $G$ at a point of inflection, where the derivative $dV_g/dG$ shown in Figure 2 (b) reaches a maximum. As the voltage on the centre finger gate $V_F$ is made more negative the step becomes narrower and eventually transforms into a point of inflection. The reason for the observed smearing of the conductance steps lies in the decrease of the ratio $\omega_y/\omega_x$ of the transverse and longitudinal frequencies that define the shape of saddle potential in the constriction [6]. Figure 2 (b) shows that $G_* = 2e^2/h$ at $V_F = 0$ and $G_* = 0.97 \times 2e^2/h$ at a large negative voltage $V_F = -1.6$ V. In the voltage range $-1.6\,\text{V} < V_F < 0$ a continuous transition between these two values is observed. We note that the exact conductance quantisation of $G_* = 2e^2/h$ at small negative $V_F$ confirms the validity of the four terminal measurement set-up. Nonlocality of resistance in this set-up can complicate measurements of the conductance of the constriction. But by deliberately ensuring that the resistance of the potentiometric contacts is much smaller than that of the single-mode channel one can neglect this nonlocality.

It is also worth mentioning that the gradual decrease of the height of the first conductance step has been detected in two-terminal conductance measurements, where $R_s$ has been subtracted. This decrease has been shown to be reproducible for different cool-downs, indicating that it cannot be explained by residual disorder scattering. In addition the lack of resonant features and the pronounced conductance steps demonstrate that we have a clean 1D system in which impurity scattering is negligible. Therefore we can exclude the possibility of "dirty Luttinger liquid" behaviour in our case. Moreover, in contrast to previous observations [7–10] where the deviation from exact quantisation increased with the length of the quantum wire, in our case we observe an increase in the deviation as the 1D region shortens. Therefore the deviation we observe must be of a different physical origin.

Let us now see if the observed decrease of the conductance steps can be explained in the model of quasi-1D ballistic transport for an adiabatic constriction. The transverse quantisation of electrons allows us to view the scattering problem in terms of the local transverse modes. The conductance is then given by the Landauer formula: $G = \frac{2e^2}{h}\sum_{kj}|t_{kj}|^2$, where $|t_{kj}|^2$ is the probability that the flux input in subband $j$ is traveling forward into subband $k$. The solution of the transverse Schrödinger equation at each value $x$ along the constriction gives energies of 1D subbands $E_n(x)$. These subbands $E_n(x)$ have the shape of smooth potential barriers offset from each other by transverse quantum $\hbar\omega_{yn}(x)$. If $\hbar\omega_{yn}$ changes slowly along $x$ then intersubband scattering is suppressed and the total transmission coefficient is equal to the sum of contributions of all subbands [5]: $T = \sum_n T_n$, $T_n = |t_{nn}|^2$. At $G \approx 2e^2/h$ the conductance can be expressed in the form

$$G = \frac{2e^2}{h}(1 - R_1 + T_2), \qquad (1)$$

where $R_1$ is the reflection coefficient of the first subband and $T_2$ — is the transmission coefficient through the second subband. When the plateaux of $G(V_g)$ appear smeared out, their height is determined by the interplay of two basic phenomena of quantum mechanics — above-barrier reflection and tunneling. If $R_1(V_g)$ reduces slower than $T_2(V_g)$ increases, then at the point of inflection of the curve $G(V_g)$ the height of the steps $G_*$ can be less than the conductance quantum $2e^2/h$.

The effect is most easily seen in an approximation of a saddle point potential $U(x,y) = V_0 + \frac{1}{2}m\big(\omega_y^2 y^2 - \omega_x^2(x - x_{\max})^2\big)$, for which the transmission coefficient via the $n$th subband $E_n(x) = V_0 + \hbar\omega_y(n - 1/2) - \frac{1}{2}m\omega_x^2(x - x_{\max})^2$ has a simple analytical expression [6]: $T_n = (1 + e^{-\pi\epsilon_n})^{-1}$, where $\epsilon_n = 2\big(E - \hbar\omega_y(n - \frac{1}{2}) - V_0\big)/\hbar\omega_x$, $n = 1, 2, \ldots$.

With increasing negative voltage $V_g$ on the split gate the saddle potential goes up and becomes narrower. Modeling of the 3D electrostatics shows that the change of $\omega_x$ is small compared to the change in $\omega_y$ and $V_0$, as $V_g$ varies. So we can assume that $\omega_y$ and $V_0$ depend linearly on $V_g$, while $\omega_x$ remains constant. Then one can readily see that the second subband moves more rapidly than the first one:

$$\begin{aligned} E_1' &= \frac{dE_1}{dV_g} = \frac{dV_0}{dV_g} + \frac{1}{2}\frac{\hbar\, d\omega_y}{dV_g}, \\ E_2' &= \frac{dE_2}{dV_g} = \frac{dV_0}{dV_g} + \frac{3}{2}\frac{\hbar\, d\omega_y}{dV_g}. \end{aligned} \qquad (2)$$

Hence, at the minimum of $dT/dV_g$ it holds that $R_1 > T_2$.

Indeed, at $T \approx 1$ the coefficients can be approximated as

$$R_1 = 1 - T_1 \approx \exp\left(-2\pi\frac{E - E_1(x_{\max})}{\hbar\omega_x}\right), \qquad (3)$$
$$T_2 \approx \exp\left(-2\pi\frac{E_2(x_{\max}) - E}{\hbar\omega_x}\right).$$

At the point of inflection, $d^2T/dV_g^2 = 0$, we have $T_2 \approx R_1(E_1'/E_2')^2$, $E = V_0 + \hbar\omega_y + \hbar\omega_x \ln(E_1'/E_2')/2\pi$, and

$$T_* \approx 1 - e^{-\pi\omega_y/\omega_x}\left(\frac{E_2'}{E_1'} - \frac{E_1'}{E_2'}\right). \qquad (4)$$

Let us estimate the deviation $\Delta = 1 - T_*$. If the steps on the plot $T(V_g)$ are to be observed, it requires that $\omega_y/\omega_x > 1.25$ [6]. Thus the exponent in (4) does not exceed 2%. On the other hand from computation of the electrostatics (see below) $dV_0/dV_g \sim \frac{1}{2}\hbar\, d\omega_y/dV_g$ and $E_2'/E_1' \approx 2$. Therefore, the conductance step can decrease by 3%. When the negative voltage on the centre finger gate approaches zero, $\omega_x$ decreases, and the effect vanishes exponentially. Thus, the simple equation (4) predicts behaviour that qualitatively agrees with our experimental results (Fig. 2).

Notice that the effect is absent if $E_2' = E_1'$, i.e. when $V_0$ has a linear dependence on the gate voltage, $V_g$ and frequencies $\omega_y$ and $\omega_x$ remain constant [6]. Essentially this is the same as if energy changes within the saddle potential are kept constant. In this case the quantisation is exact, $T = n$, because at the inflection points $E = V_0 + n\hbar\omega_y$ the reflection coefficient $R_n$ is equal to the transmission coefficient $T_{n+1}$. However, if the electrostatic potential deforms with varying gate voltage, then $G(V_g)$ dependencies for fixed Fermi energy and $G(E)$ for a fixed potential can be qualitatively different.

Realistic modelling of the electrostatics of the devices allows us to check the validity of this prediction without making any assumptions about the shape of the potential profile. The basis of the computations is given by the structural data: thickness and material of the layers, concentration of doping impurities, and three-dimensional geometry of the gates. The electrostatic potential, $U(x, y, z)$ was calculated self-consistently along with the electron density $\rho(x, y, z)$ and the statistical exchange-correlation potential $U_{\text{exc}}(x, y, z) = U_{\text{exc}}(\rho)$ [13] in the GaAlAs/GaAs/GaAlAs quantum well. The function $U_{\text{exc}}(\rho)$ was chosen to be the same as that used recently for modeling a 2DEG [14] and a 1D quantum wire [15]. It is important that exchange and correlation is taken into account, particularly at low electron densities [16].

In the middle part of the channel, $\rho$ was computed using a quasiclassical approximation from the 1D density of states, and the temperature was chosen to imitate tunneling through the potential barrier along $x$. We have found that the threshold voltages for the finger gates are different in the modelling of the electrostatics of the structures and those found experimentally. However we note that experimentally these voltages are sensitive, for example, to the charge in PMMA layer. As a result we chose the gate voltages to fit computed curves $G(V_g)$ to the experimental curves. The conductance was determined by solving the multiple-mode transmission problem with an effective potential $U_{\text{eff}}(x, y)$ in the plane of 2DEG calculated using 3D-electrostatic modelling for a given $V_g$, $V_F$ [17]. We considered 15 transverse modes in the calculations. As a result if nonadiabaticity of the constriction is responsible for the reduction in quantised values, then we would expect to find evidence for intersubband scattering. In fact intersubband mixing is absent for the modelled constrictions and transport is adiabatic.

We also studied the effect of the transverse finger gates on the electrostatic potential in the constriction. When the voltage on the outermost finger gates is equal to zero and the negative voltage on the centre finger gate is large, a potential of a triangular shape is formed along the channel, with a smoothed top and the base width defined by the separation between the outermost finger gates (solid line in Fig. 3 (a)). If there were only one finger gate in the centre then the longitudinal potential would resemble a triangle with a wider base (dotted line). Finally in the case where there are no overlaying finger gates the potential would look like a smooth wide hump (dashed line in Fig. 3 (a)). The presence of the finger gates increases $\omega_x$ and in addition decreases $\omega_y$, permiting the ratio $\omega_y/\omega_x$ to approach 1.

Figure 3 (b) shows the computed one-dimensional subbands $E_{1,2}(x)$ for $V_g = -1.47$ V and $V_F = -2.1$ V (in this case $T(E_F) = T_1 + T_2 = 0.96$). The Fermi level $E_F = 0$ is located between the tops of curves $E_1(x)$ and $E_2(x)$. Interestingly, the quantisation of the conductance $G(E)$ is exact for computed subbands $E_{1,2}(x)$.

The computed $G(V_g)$ is shown in Fig. 3 (c) together with the corresponding experimental curve. The computation was done as follows. The two first subbands $E_1(x)$ and $E_2(x)$ were shifted with changing gate voltage with velocities $dE_1/dV_g = -4.8$ meV/V and $dE_2/dV_g = -11$ meV/V ($E_2'/E_1' = 2.3$). The values of these velocities were determined from the electrostatics at adjacent values of gate voltage. The transmission coefficient was computed as the sum of transmission coefficients of the first and second one-dimensional subbands. By computing $G(V_g)$ in such a way there is no need to compute the electrostatic potential for every value of $V_g$. As a result we find that near the conductance quantum we get results very close to the more cumbersome and time-consuming computations of 3D electrostatics followed by 2D transmission. From Fig. 3 (c) we see that the predicted theoretical curve is very close to the experimental curve. The same can also be said about the curves $dG(V_g)/dV_g$. The computations demonstrate that the quantisation steps are smeared out as they are in the experiment and reproduce, with 1% precision, the observed reduction in the height of the quantised conductance steps.

In conclusion, we have demonstrated that it is possible to observe deviations from exact conductance quantisation in short adiabatic constrictions. In our devices we are able to control the potential profile in the 1D constriction using overlaying finger gates and find a gradual decrease of the height of the first conductance step from $2e^2/h$ to $0.97 \times 2e^2/h$ as we increasing apply a negative voltage to an overlaying finger gate. The effect can be explained using a saddle-point potential model taking into account the deformation of the potential with varying split-gate voltage, $V_g$. Here we find that the dependency of conductance on gate voltage is qualitatively different to that of energy such that the deviation from exact quantisation appears only in $G(V_g)$. We find that the top of the first subband moves two times slower than the second subband in response to the change of the split-gate voltage. This leads to a slower decrease of the reflection compared to simultaneous increase in the tunneling, such that when the conductance step height is smeared out it is also reduced. Realistic numerical modelling of the electrostatics of the devices and transmission through the constriction produces results that are very similar to our experimental data.

This work was funded by the NSC and the MOE, Taiwan. The work at Cambridge was supported by the EPSRC and the Royal Society, UK. The work at Novosibirsk was supported by the program "Low-dimensional quantum structures" of the Russian Academy of Science. We are grateful to A. O. Govorov, A. G. Pogosov, Kvon Ze Don, and M. V. Éntin for fruitful discussions.

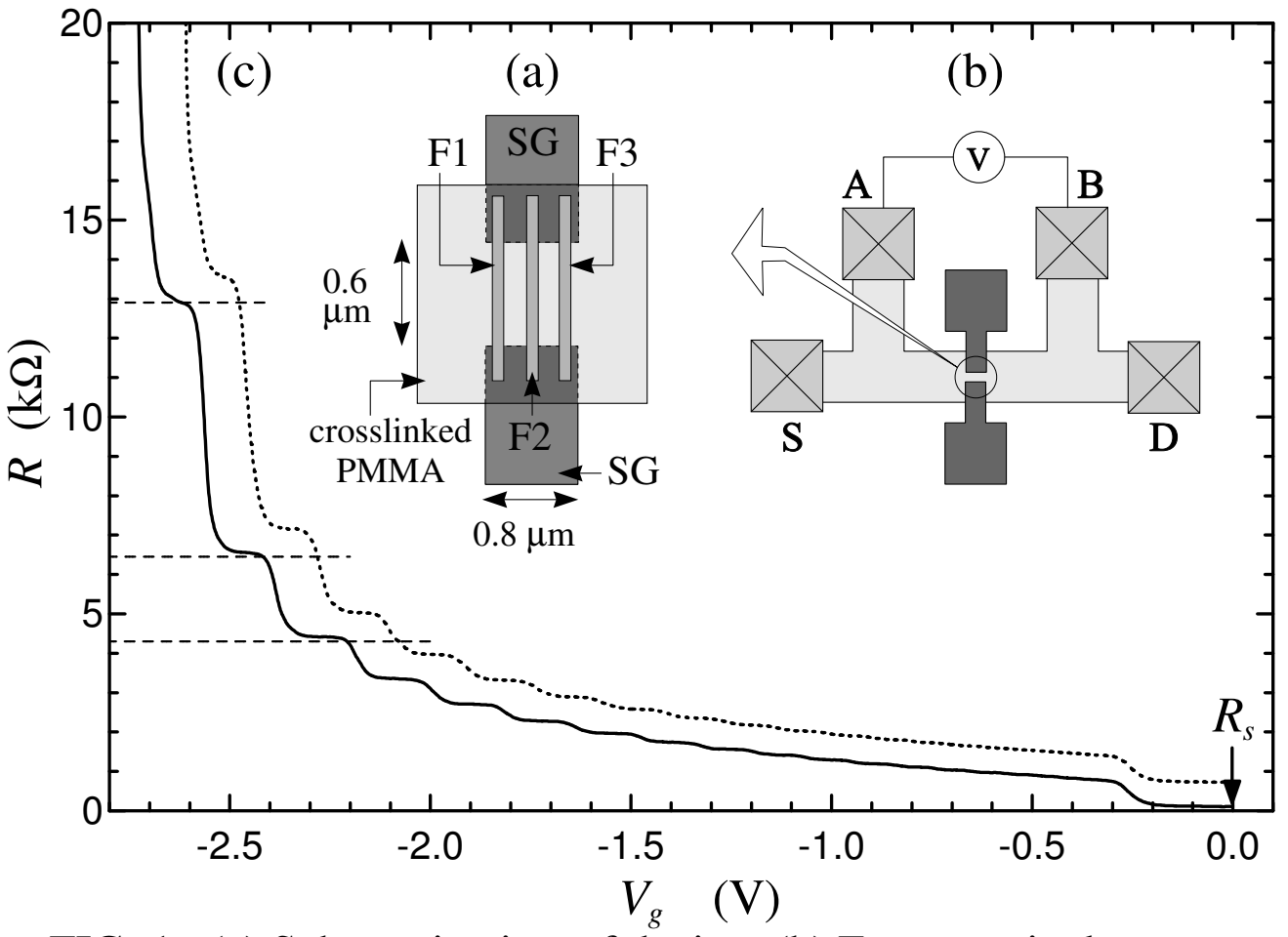

FIG. 1. (a) Schematic view of device. (b) Four-terminal measurement setup. (c) Measured dependences of resistance $R$ on the split gate voltage $V_g$, with all finger gates grounded to the 2DEG (the solid line). Horizontal dashed lines show values $R_n = h/2ne^2$ for $n = 1, 2, 3$. For comparison two-terminal measurement of $R(V_g)$ is shown by the dotted line.

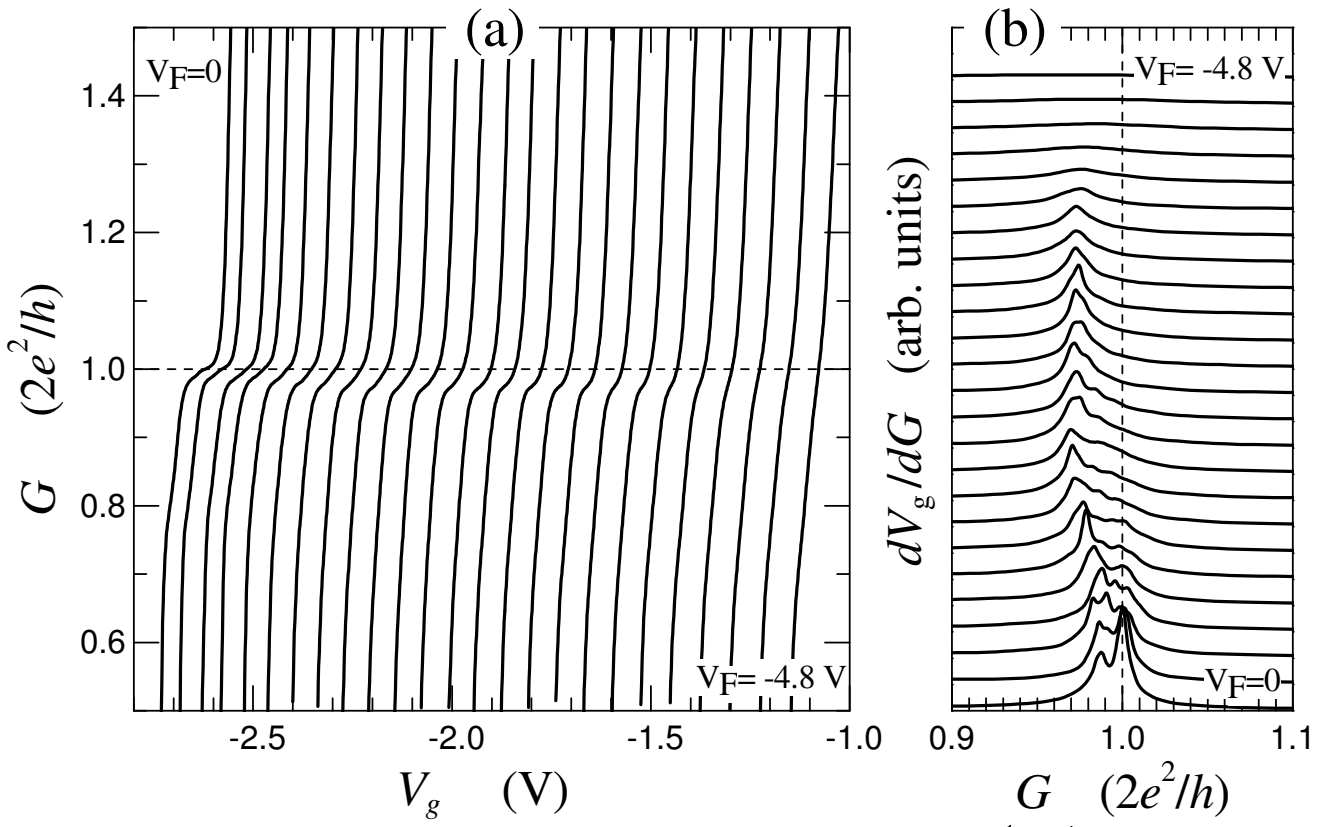

FIG. 2. (a) Measured 4-terminal conductance $G(V_g)$ at different center finger gate voltage, $V_F$. From left to right: $V_F = 0$ to $-4.8$ V in 0.2 V steps. (b) Derivative $dV_g(G)/dG$ at different center finger gate voltages, $V_F$. From bottom to top: $V_F = 0$ to $-4.8$ V in 0.2 V steps.

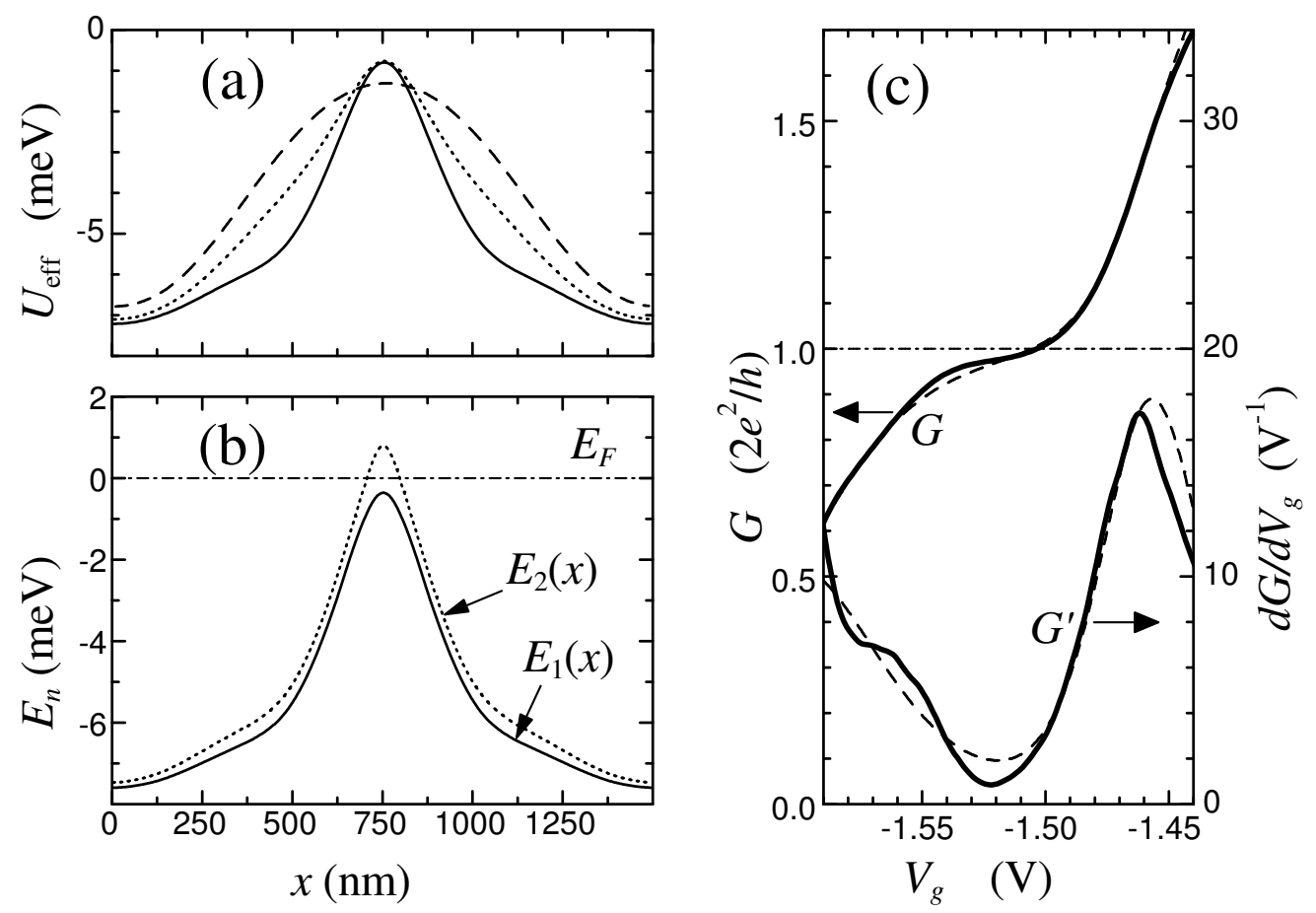

FIG. 3. (a) Effective potential $U_{\rm eff}(x, y = 0)$ computed for structures without finger gates (dashed line), with one center finger gate (dotted line), and with three finger gates (solid line). The gate voltages were chosen to adjust the transmission coefficient to unity. (b) 1D subbands $E_{1,2}(x)$ for a case characterized by surface density of charge in PMMA layer $n_i = 10^{11}\,\mathrm{cm}^{-2}$ and gate voltages $V_g = -1.47\,\mathrm{V}$ and $V_F = -2.1\,\mathrm{V}$. (c) Experimental and theoretical curves of $G(V_g)$, $dG/dV_g$ versus $V_g$. The computation (dashed line) was performed for the 1D subbands shown in (b). Measured curves (solid) refer to $V_F = -3.6$V. Theoretical curves were shifted by $-0.055$ V for better alignment.